\documentclass[journal]{IEEEtran}
\usepackage{amsfonts,subfigure,multicol,color,verbatim,
graphicx,cite,epsfig,amssymb,amsmath,cases,bm,algorithm,
algorithmic,xcolor,multirow}

\begin{document}
\markboth{IEEE Access, vol. 2016, SI on Recent Advances in Socially-aware Mobile Networking} {Peng: D2D \ldots}

\title{Recent Advances in Fog Radio Access Networks: Performance Analysis and Radio Resource Allocation}

\author{\normalsize
Mugen~Peng,~\IEEEmembership{Senior Member,~IEEE} and Kecheng~Zhang
\thanks{Mugen~Peng (e-mail: {\tt pmg@bupt.edu.cn}) and Kecheng~Zhang (e-mail: {\tt buptzkc@163.com}) are with the Key Laboratory of Universal
Wireless Communications for Ministry of Education, Beijing University of Posts and Telecommunications (BUPT), China.} }

\date{\today}
\maketitle

\begin{abstract}
As a promising paradigm for the fifth generation wireless communication (5G) system, the fog radio access network (F-RAN) has been proposed as an advanced socially-aware mobile networking architecture to provide high spectral efficiency (SE) while maintaining high energy efficiency (EE) and low latency. Recent advents are advocated to the performance analysis and radio resource allocation, both of which are fundamental issues to make F-RANs successfully rollout. This article comprehensively summarizes the recent advances of the performance analysis and radio resource allocation in F-RANs. Particularly, the advanced edge cache and adaptive model selection schemes are presented to improve SE and EE under maintaining a low latency level. The radio resource allocation strategies to optimize SE and EE in F-RANs are respectively proposed. A few open issues in terms of the F-RAN based 5G architecture and the social-awareness technique are identified as well.
\end{abstract}


\section{Introduction}

As the mobile traffic explosively grows with multitude of mobile social applications, the socially-aware mobile networks have emerged as a promising direction to exploit the people¡¯s behaviors and interactions in the social domain\textcolor[rgb]{1.00,0.00,0.00}{\cite{IOT}}. Socially-aware mobile network designs can improve shared spectrum access, cooperative spectrum sensing and device-to-device (D2D) communications, and have potential to achieve substantial gains in spectral efficiency (SE), which can meet some performance requirements of the fifth generation wireless communication (5G) systems\textcolor[rgb]{1.00,0.00,0.00}{\cite{0}}. Among all presented socially-aware mobile networking architectures, the cloud radio access networks (C-RANs) can be regarded as a promising paradigm for improving SE and energy efficiency (EE) for 5G\textcolor[rgb]{1.00,0.00,0.00}{\cite{1}}. In C-RANs, the baseband processing is centralized in the base band unit (BBU) pool and the densely deployed remote radio heads (RRHs) connected to BBU pool via fronthaul fulfill the user-centric capability. However, the constrained fronthaul with limited capacity and long time delay degrades SE and EE performances. Meanwhile, the full-centralized architecture brings heavy burdens on the computing capability in the BBU pool\textcolor[rgb]{1.00,0.00,0.00}{\cite{richard}}.

Taking full advantage of fog computing and C-RANs, fog radio access networks (F-RANs) have been proposed to tackle these aforementioned disadvantages of C-RANs as an advanced socially-aware mobile networking architecture in 5G systems\textcolor[rgb]{1.00,0.00,0.00}{\cite{2}}. In F-RANs, the RRH and the user equipment (UE) are capable of local signal processing, cooperative radio resource management, and distributed storing. Such RRHs and UEs are named the edge computing access point (EC-AP) and the edge computing user equipment (EC-UE), respectively. Since a part of UEs no longer access to the BBU pool, the heavy burdens on both fronthaul and BBU pool are alleviated, and the latency is reduced significantly as well. Therefore, F-RANs can achieve high SE/EE, low latency, and fantastic reliability for different IoT applications such as mobile vehicular connectivity, smart home, smart education, wearable healthcare devices, and industrial automation\textcolor[rgb]{1.00,0.00,0.00}{\cite{2}}.

Recently, the functionalities and technologies of F-RANs have been discussed in the 3rd generation partnership project (3GPP) standard, such as the edge cache and the system architecture requirements for fulfilling 5G systems\textcolor[rgb]{1.00,0.00,0.00}{\cite{3GPP}}. Meanwhile, many works have been done for the IoT in the F-RAN based 5G systems. The system architecture and key techniques of F-RANs have been proposed in \textcolor[rgb]{1.00,0.00,0.00}{\cite{2}}. Following, the harmonization of C-RANs and F-RANs from various points of view has been compared in\textcolor[rgb]{1.00,0.00,0.00}{\cite{n2}}. The joint computational and radio resources problems in F-RANs have been explored in\textcolor[rgb]{1.00,0.00,0.00}{\cite{5}}. Furthermore, effective caching strategies for F-RANs is given by\textcolor[rgb]{1.00,0.00,0.00}{\cite{n1}}, while the content caching and delivery techniques for 5G systems are discussed in\textcolor[rgb]{1.00,0.00,0.00}{\cite{3}} and\textcolor[rgb]{1.00,0.00,0.00}{\cite{6}}. All these publications have illustrated that the cache incorporation can significantly improve SE, EE and delay performances. However, the outage probability and ergodic capacity analysis in F-RANs are still not straightforward, which constrain the development of F-RANs. Besides, the distributed edge caching can alleviate the burdens on the fronthaul and BBU pool, meanwhile decreases latency through shortening the end-to-end communication distance, which results in the traditional radio resource strategies unsuitable for F-RANs. Since the performance analysis and radio resource allocation are two key issues for fulfilling the requirements of IoT applications in the F-RAN based 5G systems, this article presents a framework for these two issues, in which the recent advances are comprehensively summarized. Meanwhile, the challenges and open issues for the IoT in F-RANs are discussed as well.

This article is organized as follows. We discuss the edge cache based performance analysis and radio resource allocation in Section II and Section III, respectively. Open issues and challenges are discussed in Section IV. The conclusion is presented in Section V.

\section{Cache-based Performance Analysis}

As shown in Fig. 1, to fulfill the functions of socially-aware mobile networking, the hierarchical architecture consists of cloud center and fog layer in F-RANs. The cloud center is composed of cloud content cache, cloud BBU pool, and cloud controller. RRHs connect to the BBU pool via fronthaul, while the high power node (HPN) connects to the content cloud through backhaul. EC-AP integrates not only the front radio frequency (RF), but also the local distributed collaboration radio signal processing (CRSP), cooperative radio resource management (CRRM) and caching capabilities. By processing collaboratively among multiple adjacent EC-APs and even through the device-to-device (D2D) communication, the overload of the fronthaul links can be decreased, and the queuing and transmitting latency can be alleviated. When all CRSP and CRRM functions are shifted to the BBU pool, EC-AP is degenerated to a traditional RRH. EC-UEs denote by UEs accessing EC-APs and working in the D2D mode\textcolor[rgb]{1.00,0.00,0.00}{\cite{2}}.

\begin{figure}
\centering \vspace*{0pt}
\includegraphics[height=2.1in]{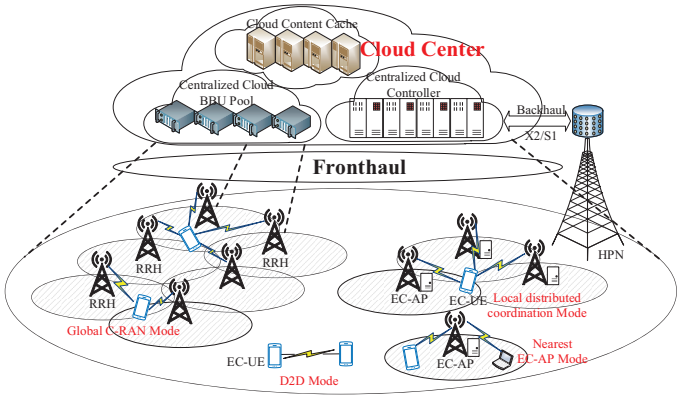}
\setlength{\belowcaptionskip}{-100pt} \caption{System architecture of the F-RAN}
\label{System}\vspace*{-10pt}
\end{figure}


\subsection{Edge caching impacting on the performance of F-RAN}

Collaborative strategy to implement caching in infrastructure and in mobile devices simultaneously is researched in\textcolor[rgb]{1.00,0.00,0.00}{\cite{6}}. With respect to sizes, the utilization is different for caching in EC-APs and caching in devices, as shown in Fig. 2 (a). For the popular social media (high rank), the best strategy is to cache it in the EC-AP. Due to the instability of opportunistic links between devices, the remaining social media with higher popularity and small size should take the chance by direct sharing via D2D. Meanwhile, different caching schemes have significant impacts on the latency, as shown in Fig. 2 (b). According to different characteristics(i.e., popularity and size), the social media contents are distributed by the infrastructure or by devices to increase the successful delivery probability, the latency is considerably reduced.

\begin{figure}
 \centering
 \subfigure[Optimal strategy of caching utilization in F-RANs\cite{6}]{
 \label{fig:subfig:a} 
 \includegraphics[width=2.4in]{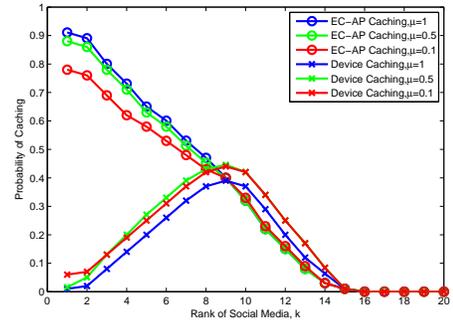}}
 \hspace{1in}
 \subfigure[Latency performance of caching utilization in F-RANs\cite{6}]{
 \label{fig:subfig:b} 
 \includegraphics[height=1.8in, width=2.4in]{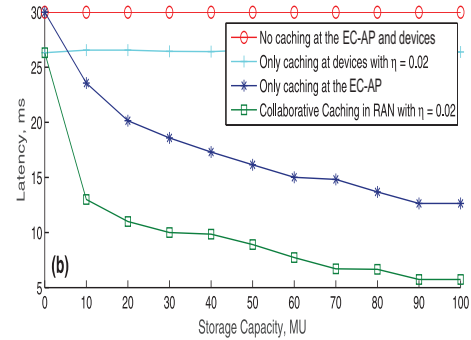}}
 \caption{Caching utilization analysis in F-RAN}
 \label{fig:subfig} 
\end{figure}
 Edge caching significantly improves SE/EE in F-RANs. In\textcolor[rgb]{1.00,0.00,0.00}{\cite{12}}, the problem of dynamic content-centric EC-AP clustering and multicast beamforming with respect to both channel condition and caching status is studied. Since each UE prefer to access the EC-APs which cache the content they want, the EC-AP clustering is influenced significantly by the caching strategies. Threes caching strategies are proposed as baselines: popularity-aware caching, random caching and probabilistic caching. For the popularity-aware caching baseline, Each BS caches the most popular contents until its storage is full. For random caching, each EC-AP caches contents randomly with equal probabilities. For the probabilistic caching, each EC-AP caches one content with the probability of its popularity. Fig. 3 (a) shows the effect of different caching strategies. It is obvious that the popularity-aware caching baseline performs best because EC-APs are more likely to cooperate with each other to achieve high cooperation gain. Random caching is only better than the no caching strategies because EC-APs can hardly cooperate with each other. The performance of probabilistic caching is in between. Meanwhile, both the power assumed by traditional radio resource and caching are jointly considered to evaluate the energy efficiency of F-RAN in\textcolor[rgb]{1.00,0.00,0.00}{\cite{13}}. The energy consumption is defined as product of the power of keeping a content object and the content object size. Particularly, a nested coalition formation game-based algorithm is proposed to find the optimal cell association scheme. Besides, a suboptimal algorithm is designed to reduce the computational complexity. Fig. 3 (b) illustrates the performance comparison of these two algorithms. The EE increases as the size of local cluster cache is enlarged, which indicates that the energy efficiency of the F-RAN outperforms the C-RAN due to the edge caching technique.

\begin{figure}
 \centering
 \subfigure[The impact of different caching techniques on SE\cite{12}]{
 \label{fig:subfig:a} 
 \includegraphics[width=2.5in]{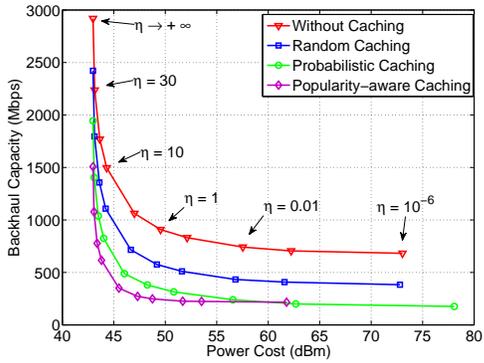}}
 \hspace{1in}
 \subfigure[The impact of different caching technique on EE\cite{13}]{
 \label{fig:subfig:b} 
 \includegraphics[width=2.8in]{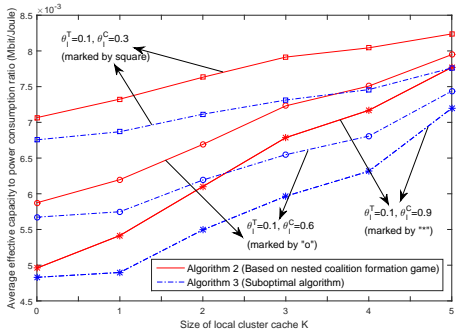}}
 \caption{Impact of caching technique to SE/EE in F-RANs}
 \label{fig:subfig} 
\end{figure}

\subsection{Mode Selection impacting on the performance of F-RAN}

In traditional C-RANs, UEs can only access the RRH to get the content that they need, which indicates only the global C-RAN mode is available in C-RANs. However, in F-RANs, due to the existing of edge caching equipment, UEs can access to D2D UEs, EC-APs or RRHs according to different channel and cache conditions to be served. Four basic access mode is shown in Fig. 1. D2D mode is enabled when the distance between two adjacent UEs with edge cache is sufficiently close and the transmitter caches the content that the receiver needs. EC-AP is triggered when no neighboring D2D UEs can provide the desired contents while the neighbour EC-AP can. Furthermore, if more than one EC-AP can provide UEs with the desired contents, the local distributed coordination mode is triggered. If the desired contents are just stored in the center edge, and the desired UE has to access all potential RRHs for the service, and hence the global C-RAN mode is triggered.

Generally speaking, one UE prefers to being served not via backhaul or fronthaul, but through accessing the local D2D UEs or EC-APs which cache the desired content to achieve high SE/EE and low latency. The major reasons include: 1) the nearer D2D UE or EC-AP brings higher signal to interference plus noise ratio (SINR), which indicates high SE and EE; 2) the D2D UE or EC-AP usually consumes less energy and suffer lower latency because the contents are delivered to UEs locally without passing through the fronthaul or backhaul. Compared with the pure C-RAN mode, in which all contents are achieved by the central cache server, the adaptive model selection in F-RANs can significantly improve SE and decrease latency.

Except for the SE and latency, green communication is another key factor in F-RANs. Ref.\textcolor[rgb]{1.00,0.00,0.00}{\cite{modeselect1}} compares the energy consumption of D2D mode and cellular mode to evaluate the cost performance of different mode selection. It can be shown in Fig. 4(b) that when the distance between two D2D users is sufficiently short, the energy of the D2D mode is significant reduced because the high SINR and low transmit power due to the short communication distance, and the low circuit power of the D2D transmitter. In F-RANs, D2D is an effective access technique to provide UEs with low latency and low EE with constrained cache capacity, while the global C-RAN mode can afford all the contents from the centralized cache, which results in not good SE and EE performance. UEs should select proper association mode, which is based on the desired content in the edge cache, the desired SINR, and the energy consumption.

To illustrate the impact of different access modes, in\textcolor[rgb]{1.00,0.00,0.00}{\cite{modeselect1}}, the authors illustrate the latency performance under different access mode. Local Video base mode (local VB) means one UE can communicate with another UE locally to get the contents, which is the same with the D2D mode in F-RANa. In Sub video base (subVB) mode, one UE access more than one neighboring access points to get the contents, which is the same with the local distributed cooperation mode in F-RAN. Video base (VB) mode means one UE access the nearest access point and it is the same with nearest EC-AP mode. Outside mode means one UE has to connect the BBU pool to get the contents it wants, which is the global C-RAN mode. It can be seen from Fig. 4 (a) that the local VB mode achieves the best performance. The performance of subVB mode and the VB mode are similar while the subVB performs better because of the cooperation gain. The latency of outside mode is biggest because it wastes much time on getting contents from the Internet and transmitting them through fronthaul.

\begin{figure}
 \centering
 \subfigure[Average delay of different access mode\cite{modeselect1}]{
 \label{fig:subfig:a} 
 \includegraphics[width=2.7in]{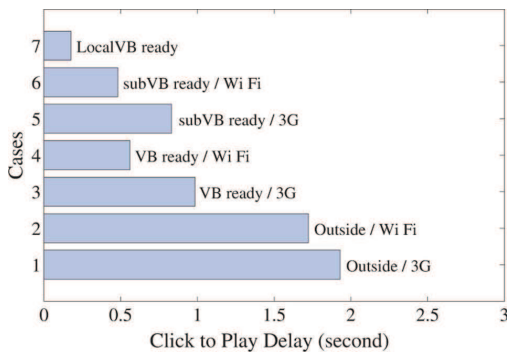}}
 \hspace{1in}
 \subfigure[Percentage of energy saved in D2D mode compared with cellular mode]{
 \label{fig:subfig:b} 
 \includegraphics[width=2.4in]{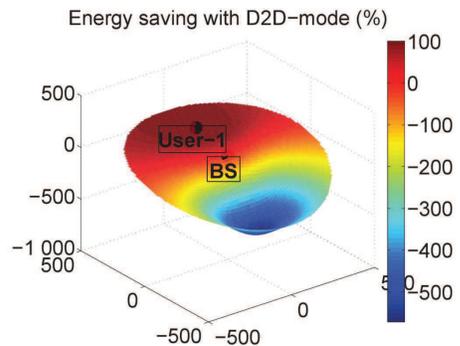}}
 \caption{Impact of access mode to the performance of F-RAN}
 \label{fig:subfig} 
\end{figure}

\section{Radio Resource Allocation}

Compared with that in C-RANs, radio resource allocation in F-RANs is more complex but advanced because the deployment of equipment with edge caching such as EC-APs and D2D UEs should be considered. Precoding design, resource block allocation, user scheduling, and cell association should be jointly designed to optimize SE, EE and latency performances. Three kinds of performances are categorized to illustrate the impact of radio resource allocation in F-RANs.

\textbf{\emph{Spectral Efficiency Optimization:}} To optimize SE in C-RANs, one UE can access to a certain subset of neighboring RRHs. Such subsets of neighboring RRHs are mainly determined by disjoint clustering scheme or user-centric clustering scheme\textcolor[rgb]{1.00,0.00,0.00}{\cite{5}}, in which the cluster formation is based on the reference signal received power threshold. Disjoint clustering scheme can mitigate the inter-cell interference but the edge users may suffer serious interference from neighboring cluster. In user-centric clustering scheme, there exists no edge users, thus the inter-cluster interference decreases. Besides the large-scale CRSP, C-RAN executes power allocation and dynamical scheduling to enhance SE. However, in F-RAN, cell association is quite different from that in C-RAN because users prefer to access EC-APs which cache the contents they interest in. The serving cluster can be determined by two conditions: 1)The EC-AP caches contents the UE wants; 2) The reference signal received power from the EC-AP to the UE is high enough. Once the serving cluster is formulated, user scheduling and power allocation is executed to optimize the spectral efficiency of the network. Besides, in F-RAN, spectral efficiency is no longer an important goal to pursuit because high spectral efficiency means heavy burden on the fronthaul, whose capacity may be not high enough to afford too much data traffic. Thus, the tradeoff between the fronthaul capacity and the transmit power should be highlighted.

To illustrate the impact of the cell association in F-RANs, in\textcolor[rgb]{1.00,0.00,0.00}{\cite{11}}, the ergodic rate for both EC-AP UEs and device-to-device UEs by taking into account the different nodes locations, cache sizes as well as user access modes are analyzed and evaluated. As shown in Fig. 4(a), the numerical results of the ergodic rate for the local distributed coordination mode with different EC-AP cache sizes versus cluster radius threshold $L_c$ are validated. It illustrates how ergodic rate varies with the number of EC-APs in the cluster when the EC-AP cache size is fixed. It suggests that more EC-APs serve the desired UE with the increase of signal and the decline of interference and leads to the enlargement of cluster SIR and results in the improvement of ergodic rate of local distributed coordination mode. Similarly, the larger cache size of EC-APs $C_f$ suggests that there are more opportunities for the desired UE to get the contents that it needs, which leads to a higher SE performance.

\begin{figure}
 \centering
 \subfigure[Ergodic rate versus cluster radius\cite{11}]{
 \label{fig:subfig:a} 
 \includegraphics[width=2.5in]{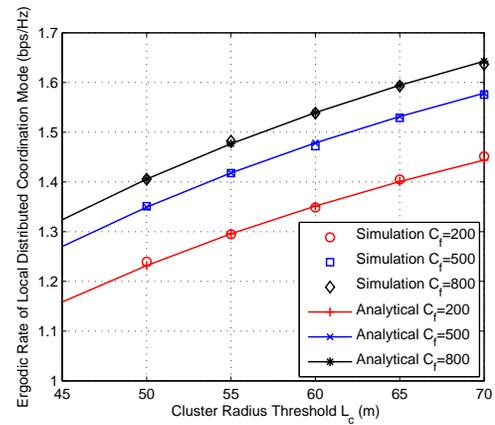}}
 \hspace{1in}
 \subfigure[Average EE vs. size of cluster content cache\cite{13}]{
 \label{fig:subfig:b} 
 \includegraphics[width=2.6in]{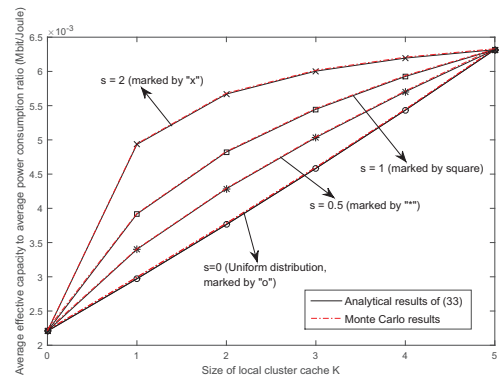}}
 \caption{Impact of cache based cell association to SE/EE performance of F-RAN}
 \label{fig:subfig} 
\end{figure}

\textbf{\emph{Energy Efficiency Optimization:}} Since C-RAN is a fully centralized network, RRHs only operate as soft relay by compressing and forwarding the received signals from UEs to the BBU pool and all of the signal and baseband processing are performed in the BBU pool. All desired contents that UEs wants are stored in the BBU pool and delivered to UEs by fronthaul, which consumes much energy and results in the low EE. The EE optimization in C-RANs focuses on maximizing the total overall SE and minimizing the energy consumed in the BBU pool. However, in F-RANs, the EC-APs are capable of caching contents and executing signal and baseband processing. Thus, some UEs can get the preferred contents from local EC-APs without through fronthaul and backhaul. With more contents cached in local EC-APs, the EE performance of F-RAN becomes better. The EE optimization in F-RANs should highlight the effect of local caching. In\textcolor[rgb]{1.00,0.00,0.00}{\cite{13}}, the numerical results of the average effective EE versus the size of cluster content cache are validate, as shown in Fig. 4(b). With the larger size of local cluster cache, the EE performance of F-RANs increases almost 3 times because more UEs can access EC-APs to locally get the contents cached.

\textbf{\emph{Latency optimization:}} In C-RANs, all signals and contents are centralized to the BBU pool from RRHs through fronthaul. Since RRHs have no capability of signal processing and storage, the latency between RRHs and UEs is much smaller than that between RRHs and the BBU pool. The latency optimization problem in C-RANs focus on minimizing the latency of fronthaul, which determines the whole latency of the network. However, in F-RANs, the latency optimization problem is more complex. EC-APs can provide contents that UEs need, which results in the significant decrease of the latency of fronthaul, while the latency from EC-APs to UEs can not be ignored. To illustrate the impact of the cache placement to the latency, \textcolor[rgb]{1.00,0.00,0.00}{\cite{15}} formulates an average download delay minimization problem subject to the caching capacity constraint of each base station. Fig. 5 shows that the results given by the efficient radio resource allocation algorithm almost approaches the optimal value that obtained by the exhaustive search. Besides, when the cache size becomes enlarged, the download delay decreases dramatically. Therefore, it is anticipated that the local caching technique is promising to improve the delay performance of F-RANs.
\begin{figure}
\centering \vspace*{0pt}
\includegraphics[height=2.3 in]{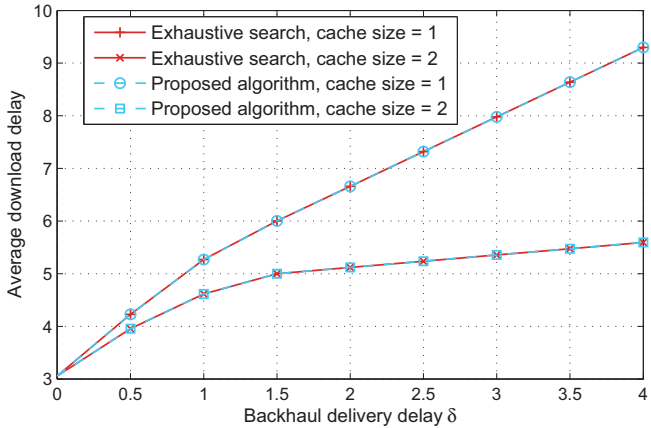}
\setlength{\belowcaptionskip}{-100pt} \caption{Average download delay versus backhaul delivery delay\cite{15}}
\label{System}\vspace*{-10pt}
\end{figure}

\section{Open Issues and Challenges }

Although the performance analysis and radio resource allocation for F-RANs have been researched, there are still many challenges and open issues remaining to be discussed in the future, including the F-RAN architecture for 5G, and the social-awareness F-RAN.

\subsection{F-RAN architecture for 5G}

It is envisioned that 5G will bring a 1000x increase in terms of area capacity compared with 4G, achieve a peak rate in the range of tens of Gbps, support a roundtrip latency of about 1 ms as well as connections for a trillion of devices, and guarantee ultra-reliability. Since F-RANs are evolved from the existing HetNets and C-RANs, it is fully compatible with the other 5G systems. The advanced 5G techniques, such as the massive multiple-input multiple-output, cognitive radio, millimeter wave communications, and non-orthogonal multiple access, can be used directly in F-RANs. There are some apparent advantages in F-RANs, including the local caching, real-time CRSP, and flexible CRRM at the edge devices, the rapid and affordable scaling that make F-RANs adaptive to the dynamic traffic and radio environment, and low burdens on the fronthaul and the BBU pool. Furthermore, the different IoT applications in 5G, such as
mobile vehicular connectivity, smart city, and industrial automation, should be based on the F-RAN. It can be anticipated that F-RANs will be qualified for meeting the high SE, EE, low latency and high reliability for different kinds of IoTs in 5G.

\subsection{Social-awareness F-RAN}

In the traditional D2D scenario, one UE is able to connect to any other UE via D2D once the channel gain between them is high enough. However, in practical, UEs are usually carried by human beings who are actively involved in social interactions and the social relationships of UEs are quite different from each others. One UE usually does not communicate with another unfamiliar UE due to the security. According to the locations, interests and background, UEs can be divided into different communities. UEs in the same community tend to exchange contents with each other while they seldom connect to UEs in other communities. Compared with the traditional D2D scenario, this social feature becomes more apparent in F-RANs as more contents are cached in the edge, and UEs prefer to be associated with the edge equipment, such as EC-APs and D2D UEs, other than the centralized BBU pool. Therefore, the social relationship affects the success probability of a D2D communication, which contributes a lot to the SE/EE and latency in F-RANs. Unfortunately, the performance analysis and radio resource allocation seldom take the social relationship into consideration and how social relationship effects the performance analysis and radio resource allocation in F-RANs is still not straightforward. Thus, the social aware F-RAN, which leverages the community property of social networks with the communication systems, is worthy to be researched.

\section{Conclusion}

This article has outlined and surveyed the recent advances of the performance analysis and radio resource allocation for the socially-aware mobile networking in fog radio access networks (F-RANs). Compared with the typical cloud radio access networks (C-RANs), the performance analysis and radio resource allocation in F-RANs are more complex but advanced due to the local edge cache and adaptive model selection. The spectral efficiency (SE), energy efficiency (EE), and latency of F-RANs as three major issues have been exploited in this article. In particular, to show how different caching and model selection strategies impact on SE, EE and latency, some impressive numerical results have been summarized. Nevertheless, given the relative infancy of the field for the F-RAN based 5G system, there are still quite a number of outstanding problems that need further investigation. Notably, it is concluded that greater attention should be focused on the F-RAN architecture and the social-awareness F-RAN. The presented performance analysis and radio resource allocation in F-RANs provide breakthroughs of theories and technologies for the advanced 5G systems, and it is anticipated to transfer these two key issues of F-RANs that discussed in this article to the standards organizations.

\begin{IEEEbiography}[{\includegraphics[width=1in]{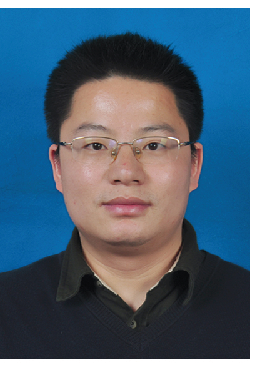}}]{Mugen Peng}
(M'05--SM'11) received the B.E. degree in electronics engineering from the Nanjing University of Posts and Telecommunications, Nanjing, China, in 2000, and the Ph.D. degree in communication and information systems from the Beijing University of Posts and Telecommunications (BUPT), Beijing, China, in 2005. Afterward, he joined BUPT, where he has been a Full Professor with the School of Information and Communication Engineering since 2012. In 2014, he was an Academic Visiting Fellow with Princeton University, Princeton, NJ, USA. He leads a Research Group focusing on wireless transmission and networking technologies with the Key Laboratory of Universal Wireless Communications (Ministry of Education), BUPT. His main research areas include wireless communication theory, radio signal processing, and convex optimizations, with a particular interests in cooperative communication, self-organization networking, heterogeneous networking, cloud communication, and internet of things. He has authored/coauthored over 60 refereed IEEE journal papers and over 200 conference proceeding papers.

Dr. Peng was a recipient of the 2014 IEEE ComSoc AP Outstanding Young Researcher Award, and the best paper award in IEEE WCNC 2015, WASA 2015, GameNets 2014, IEEE CIT 2014, ICCTA 2011, IC-BNMT 2010, and IET CCWMC 2009. He received the First Grade Award of the Technological Invention Award in the Ministry of Education of China for the hierarchical cooperative communication theory and technologies, and the First Grade Award of Technological Invention Award from the China Institute of Communications for the contributions to the self-organizing networking technology in heterogeneous networks. He is on the Editorial/Associate Editorial Board of the \emph{IEEE Communications Magazine}, \emph{IEEE Access}, \emph{IET Communications},
\emph{International Journal of Antennas and Propagation (IJAP)},
and \emph{China Communications}. He has been the guest leading editor
for the special issues in the \emph{IEEE Wireless Communications}, \emph{IEEE Access}, and \emph{IET Communications}.
\end{IEEEbiography}

\begin{IEEEbiography}[{\includegraphics[width=1in]{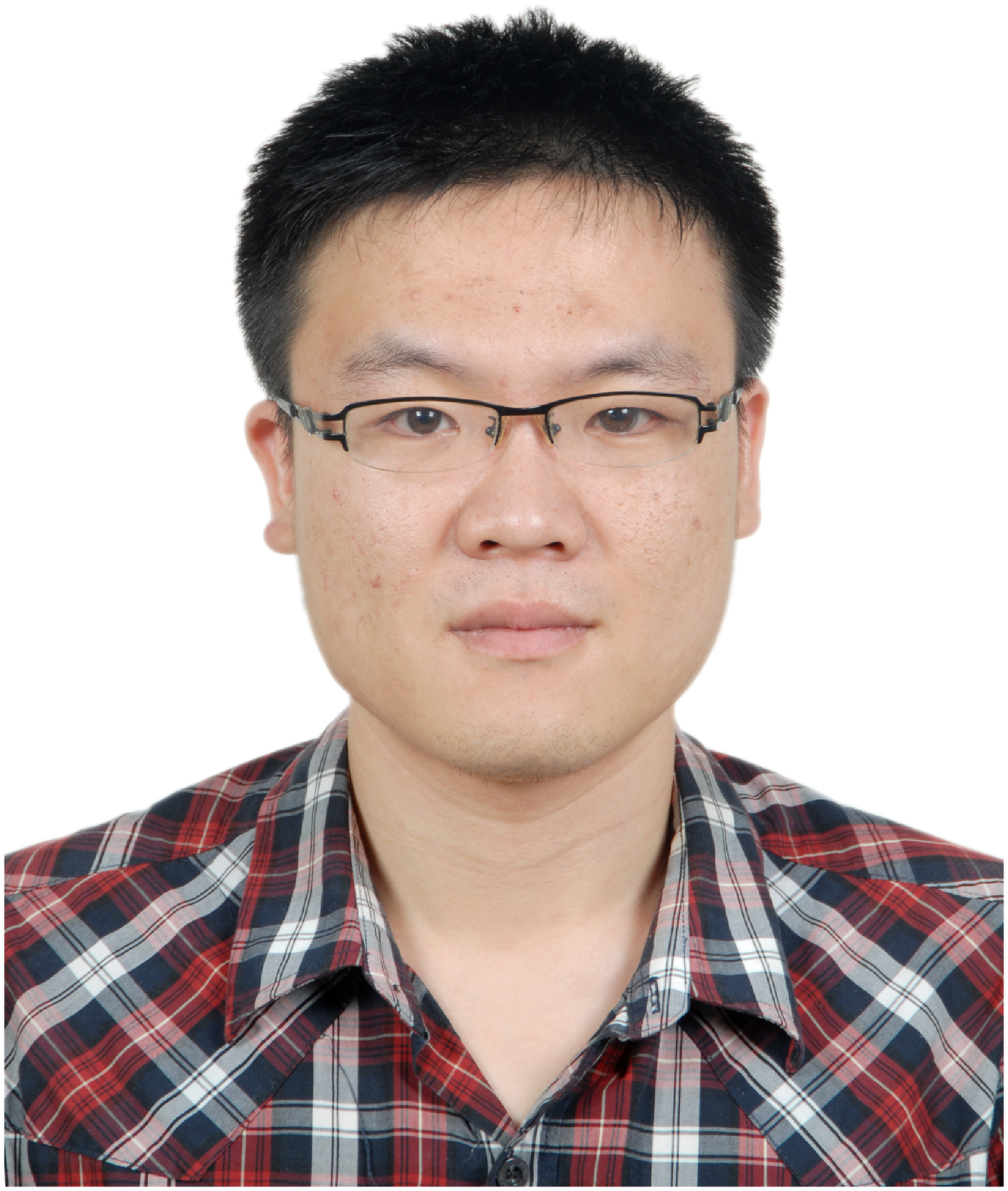}}]{Kecheng Zhang}
received the B.S. degree in telecommunication engineering from
Beijing University of Posts and Communications (BUPT), China, in
2012. He is currently pursuing the Ph.D. degree in the Key
Laboratory of Universal Wireless Communications for Ministry of
Education at BUPT. His research interests include the radio resource
allocation optimization and cooperative communications in
heterogeneous cloud radio access networks.
\end{IEEEbiography}
\vfill

\end{document}